\documentclass{article}
\usepackage[T1]{fontenc} 
\usepackage[utf8]{inputenc} 
\usepackage{ismir,amsmath,cite,url}
\usepackage{amsfonts}
\usepackage{graphicx}
\usepackage{color}
\usepackage{arydshln}
\usepackage{booktabs}
\usepackage{multirow}
\usepackage{adjustbox}
\usepackage{tabularx}
\usepackage{caption} 
\usepackage[bookmarks=false,hidelinks]{hyperref}

\title{MuSLCAT: Multi-Scale Multi-Level Convolutional Attention Transformer for Discriminative Music Modeling}

\threeauthors
   {Kai Middlebrook} {\small University of San Francisco \\ {\tt\small kaimiddlebrook@gmail.com}}
   {Shyam Sudhakaran} {\small University of San Francisco\\ {\tt\small shyamsnair@protonmail.com}}
   {David Guy Brizan} {\small University of San Francisco \\ {\tt\small dgbrizan@usfca.edu}}

\begin{document}

\maketitle
\begin{abstract}

In this work, we aim to improve the expressive capacity of waveform-based discriminative music networks by modeling both sequential (temporal) and hierarchical information in an efficient end-to-end architecture. We present MuSLCAT, or Multi-scale and Multi-level Convolutional Attention Transformer, a novel architecture for learning robust representations of complex music tags directly from raw waveform recordings. We also introduce a lightweight variant of MuSLCAT called MuSLCAN, short for Multi-scale and Multi-level Convolutional Attention Network. Both MuSLCAT and MuSLCAN model features from multiple scales and levels by integrating a frontend-backend architecture. The frontend targets different frequency ranges while modeling long-range dependencies and multi-level interactions by using two convolutional attention networks with attention-augmented convolution (AAC) blocks. The backend dynamically recalibrates multi-scale and level features extracted from the frontend by incorporating self-attention. The difference between MuSLCAT and MuSLCAN is their backend components. MuSLCAT's backed is a modified version of BERT \cite{Devlin2019BERT}. While MuSLCAN's is a simple AAC block. We validate the proposed MuSLCAT and MuSLCAN architectures by comparing them to state-of-the-art networks on four benchmark datasets for music tagging and genre recognition. Our experiments show that MuSLCAT and MuSLCAN consistently yield competitive results when compared to state-of-the-art waveform-based models, yet requires considerably fewer parameters.
\end{abstract}

\section{Introduction}\label{sec:introduction}
    \begin{figure}[ht]
    \centering
    \includegraphics[width=\linewidth]{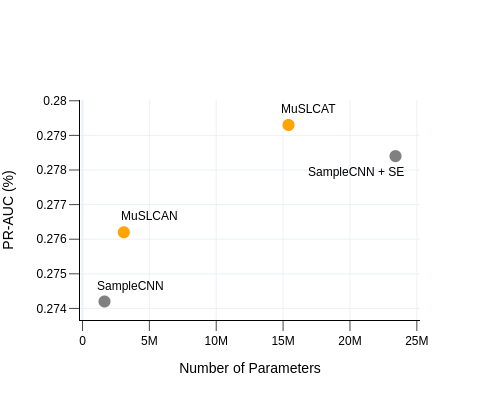}
    \caption{Music tagging performance (PR-AUC) on the MTG-Jamendo large dataset as a function of model complexity for state-of-the-art models (grey) and our proposed MuSLCAT/MuSLCAN architecture (orange). All models learn directly from raw waveform input. MuSLCAT improves performance relative to the top waveform-based network (SampleCNN + SE) yet requires fewer parameters (reduction of $34.2$\%). While MuSLCAN yields competitive results with approximately $86.8$\% fewer parameters than SampleCNN + SE.}
    \label{figure:performance_complexity}
    \end{figure}
    
    Automatic music genre recognition (MGR) aims to extract genre tag(s) from an audio recording of music. It is an extension of the well-studied task of automatic genre classification (MGC), where the aim is to extract a single genre tag from a musical piece. MGC can be considered a subtask of music auto-tagging (MAT), where the aim is to extract semantic tags (i.e., high-level descriptive keywords related to genres, moods, time period, and instruments, etc.) from a song. An accurate MGR system enables diverse applications in music production, computational musicology, and especially music search and recommendation, where genres are widely used to organize music and aid in retrieval and discovery \cite{Casey2008ContentBasedMusicRetrieval,Lamere2008SocialTaggingAndGenreChallenges, fabbri2004genretheory}. 
    
    Many recent studies on MAT, MGC, and MGR implement a convolutional neural network (CNN) architecture \cite{lee2018samplecnnAppliedSciences, kim2018samplelevelcnnIEEE, pons2017end, Choi2016AutoTaggingFCN}. Several variants of the CNN have been explored in these studies. At a high level, these architectures differ in the input representation and amount of domain knowledge leveraged.  Most previous studies convert the raw waveform of a music audio recording into a spectrogram (particularly the mel-spectrogram) input representation \cite{Pons2017DesigningMusicallyMotivated, Choi2016AutoTaggingFCN, lee2017multilevelmusictagging} and leverage popular CNN methods from Computer Vision (CV) like VGG-style \cite{simonyan2015vgg} networks (i.e., a deep stack of convolutional blocks with $3 \times 3$ 2D filters) \cite{Choi2016AutoTaggingFCN,lee2017multilevelmusictagging,pons2017end, Nam2019DeepForMusicMetaReview}. On the other hand, a small selection of recent studies experiment with systems that learn directly from the raw waveform audio inputs. The most successful waveform-based networks also adapted several approaches from CV including VGG-style \cite{simonyan2014vgg} networks, residual connections \cite{he2016ResNet}, and squeeze-and-excitation \cite{Hu2018SqueezeandExcitationN} attention. The primary differences from spectrogram-based networks and these networks are that they use 1D convolutions and avoid preprocessing the raw waveform audio.
    
    Despite the success of CNN architectures for music tagging and genre recognition tasks, the inductive biases in the convolutional layer may be too restrictive to account for the complexities of certain music tags. In particular, music genres. They can be expressed on different timescales (i.e., local vs global structure), have different levels of abstraction (e.g., language, region, sound, tempo, subject), and overlap with other genres (e.g., rock - country, hip hop - funk) \cite{lee2017multilevelmusictagging, Lamere2008SocialTaggingAndGenreChallenges, fabbri2004genretheory}. For example, identifying factors of pop music typically include repeated choruses and hooks and rhythms or tempos that can be easily danced to. It's also common for pop music to borrow elements from other genres such as rock, dance, and Latin. Yet, most successful music tagging networks operate on a single timescale and implement sample-level filters (i.e., filter size < 6) \cite{pons2017end, lee2018samplecnnAppliedSciences,kim2018samplelevelcnnIEEE}.
    
    The self-attention mechanism \cite{cheng2016selfattention} was recently proposed to model sequential data for natural language processing (NLP) tasks \cite{vaswani2017AttentionIA}. The key idea behind self-attention is to produce a weighted average of values computed from hidden units. Unlike the pooling or the convolutional operator, the weights used in the weighted average operation are produced dynamically via a similarity function between hidden units. As a result, the interaction between input signals depends on the signals themselves rather than being predetermined by their relative location like in convolutions. This enables self-attention to capture long-range interactions without increasing the number of parameters.

    In this work, we attempt to improve the expressive capacity of waveform-based discriminative music networks by modeling both sequential (temporal) and hierarchical information in an efficient end-to-end architecture. We present MuSLCAT, or Multi-scale and Multi-level Convolutional Attention Transformer, a novel architecture for learning robust representations of complex music tags directly from raw waveform recordings. We also introduce a lightweight variant of MuSLCAT called MuSLCAN, short for Multi-scale and Multi-level Convolutional Attention Network. Both MuSLCAT and MuSLCAN model features from multiple scales and levels by integrating a frontend-backend architecture. The frontend targets different frequency ranges while modeling long-range dependencies and multi-level interactions by using two convolutional attention networks with attention augmented convolution (AAC) blocks. The backend dynamically recalibrates multi-scale and level features extracted from the frontend by incorporating self-attention. The difference between MuSLCAT and MuSLCAN is their backend components. MuSLCAT's backed is a modified version of BERT \cite{Devlin2019BERT}. While MuSLCAN's is a simple AAC block. We validate both MuSLCAT and MuSLCAN by comparing them to state-of-the-art networks on four benchmark datasets for music tagging and genre recognition. Our experiments show that MuSLCAT and MuSLCAN consistently yield competitive results to state-of-the-art waveform-based models, yet requires considerably fewer parameters. 
    
    The main contributions of MuSLCAT and MuSLCAN compared to previous studies for waveform-based music tagging networks \cite{kimComparisonSampleCNNIEEE, lee2017multilevelmusictagging, lee2018samplecnnAppliedSciences} include:
        \begin{enumerate}
        \item Attending jointly to channel and temporal subspaces by implementing attention augmented convolution blocks.
        \item Emphasizing different frequencies by using two convolutional attention network branches.
        \item Leveraging all the information in multi-level or multi-scale and level features by recalibrating features using self-attention instead of summarizing them with global pooling.
        \item Modeling long-range dependencies and multi-scale and level feature interactions by integrating either AAC or a modified BERT \cite{Devlin2019BERT} in the backend. 
        \item Training in an end-to-end fashion (i.e., no pretraining or preprocessing).
        \item Reducing the number of parameters without sacrificing representational power.
        \end{enumerate}

    
    The rest of this paper is structured as follows. We review previous music tagging methods and relevant research for sequence modeling in Section \ref{sec:related_work}. In Section \ref{sec:method}, we formally describe the proposed MuSLCAT architecture. We describe our experiments in Section \ref{sec:experimentsetup} and discuss results in Section \ref{sec:results-discussion}. We offer concluding remarks and ideas for future work in Section \ref{sec:conclusion}.

\section{Related Work}\label{sec:related_work}
    In this section, we describe the most successful architectures for music tagging and classification, discuss complementary studies, and briefly introduce relevant research in natural language processing (NLP) and computer vision (CV).

    \subsection{Deep neural networks for discriminative music tasks}
        Recent research shows feature learning approaches that use a deep neural network (DNN) outperform traditional hand-crafted feature modeling approaches for solving discriminative music tasks such as MAT and MGR \cite{lee2018samplecnnAppliedSciences, kim2018samplelevelcnnIEEE, pons2017end, Choi2016AutoTaggingFCN, choi2016ConvolutionalRNN, dieleman2014end,won2020eval,won2020harmonic}. Perhaps the most ubiquitous DNN variant is the convolutional neural network (CNN).  It exhibits several implicit biases such as weight sharing and shift equivariance (i.e., translation equivariance) and can be configured to learn invariant feature representations. It also pairs well with current compute and modeling resources, which helps improve training speed and ease of implementation. Many recent CV and NLP studies introduce operations or design configurations to enhance the modeling power of the CNN and reach state-of-the-art performance as a result \cite{he2016ResNet,Hu2018SqueezeandExcitationN,hu2018gatherandexcite}. MIR studies quickly began integrating these CV and NLP CNN methods for discriminative music tasks. 
        
        We detail the state-of-the-art networks for music tagging and genre recognition in the two subsections below\footnote{All networks described in these section use batch normalization \cite{ioffe2015batchnorm} followed by ReLU activation after each convolutional operation, unless another method is explicitly stated.}. Each subsection represents a popular input representation and the networks that use it. 
        
        
        \subsubsection{Spectrogram-based networks}\label{subsec:specmodels}
            Previous studies tend to use spectrogram or one of its variants like mel-spectrogram as the input representation for a network \cite{Choi2016AutoTaggingFCN,pons2017end, lee2017multilevelmusictagging, won2019cnnsa, won2020harmonic}. This requires a preprocessing stage with multiple computational steps including Short-Time Fourier Transforms (STFT), absolute value operation, linear-to-mel mapping, and magnitude compression to transform the raw waveform audios into spectrograms before training and inference can occur. The preprocessing stage can be seen as special case of neural networks where linear transforms (STFT and linear-to-mel mapping) and non-linearity functions (absolute value operation and magnitude compression) are fixed by hand design \cite{humphrey2013featurelearning,Nam2019DeepForMusicMetaReview,kimComparisonSampleCNNIEEE}. The hyper-parameters of the "fixed networks" such as window size, hop size, and mel-filter bank size are sensitive to different audio domains, so its common to select them from the best practice in each task. We refer to models that use spectrograms as inputs as ``spectrogram-based.''
            
            A deep fully convolutional network (FCN) \cite{Choi2016AutoTaggingFCN} was one of the first deep learning approaches to reach state-of-the-art performance for music tagging. It consists of only convolutional layers without any fully-connected (FC) layers. The preprocessing step for the FCN involves converting an audio segment of 29.1 seconds to a $96 \times 1366$ mel-spectrogram. The mel-spectrogram is then fed as input to a 4-layer CNN. Each layer contains a convolutional operation with $3 \times 3$ 2D filters followed by a max-pooling operation. Different sizes of strides are used for max-pooling layers ((2, 4), (4, 5), (3, 8), (4, 8)) to increase the size of receptive fields to cover the entire input mel-spectrogram ($96 \times 1366$). 
            This training method is termed ``song-level'' training \cite{won2020eval} since most music tagging and genre recognition datasets can only provide approximately 30s-long audio clips for songs.
        
            MusiCNN \cite{pons2017end} enhanced the representational capacity of the CNN for music tagging by leveraging some music domain knowledge. In particular, the researchers encourage MusiCNN to learn timbral characteristics and temporal patterns by carefully designing vertical and horizontal filters in the first convolutional layer. The vertical filters model pitch invariant timbral features by employing $38 \times 7$ filters to capture sub-band information across a short period of time and then enforce pitch invariance by using global max-pooling across the frequency axis. The horizontal filters model temporal energy envelope of the audio by first applying global mean-pooling across the frequency axis of input mel-spectrogram and then feeding the output to long horizontal $1 \times 165$ filters.  The extracted timbral (vertical output) and temporal  (horizontal output) features are concatenated channel-wise and forwarded to a backend CNN network. The backend contains three 1D convolutional layers with residual connections \cite{he2016ResNet} followed by two FC blocks. MusiCNN also leverages multi-level feature modeling. It first performs channel-wise concatenation of features extracted from the frontend and each convolutional layer of the backend. Channel statistics are then computed by applying both global max and mean pooling respectively across the temporal dimension of each channel. The extracted max and mean channel statistics are channel-wise concatenated and fed to the two FC blocks to predict relevant tags. Different from FCN, the MusiCNN uses short 3s-long audio excerpts (which are converted to mel-spectrograms) as its inputs during training. This short audio training methods is termed "chunk-level" training \cite{won2020eval}.
            
            HarmonicCNN \cite{won2020harmonic}, like MusiCNN, leverages some music domain knowledge in its design. It takes advantage of trainable harmonic band-pass filters in the first layer, which make the model more flexible to learning representations from the spectrogram. The extracted harmonic features are fed to a 7-layer CNN. Each layer contains a  convolutional operation with sample-level $3 \times 3$ 2D filters followed by either a $2 \times 2$ or $2 \times 3$ max-pooling operation. The CNN output is then passed through two FC blocks to predict relevant tags. It also uses 5s-long audio segments (which are converted to spectrograms) to perform chunk-level training.
            
            ShortChunkCNN \cite{won2020eval} takes advantage of the popular VGG-style CNN network \cite{simonyan2015vgg} that is prevalent for CV tasks. It is a 7-layer CNN frontend and a 2-layer fully-connected backend. In the frontend, each layer contains a convolutional operation with $3 \times 3$ 2D filters followed by a max-pooling operation. The frontend was also enhanced with residual connections \cite{he2016ResNet} in the same study \cite{won2020eval}. We refer to this enhanced network as ShortChunkCNN + Res. Both networks use short 3.69s-long chunks of audio (which are converted to spectrograms) to perform chunk-level training, hence their names. The main difference between FCN and these two networks is that they use chunk-level training and smaller max-pooling operations 
        
            The sequential nature of music may lead one to believe modeling long-range dependencies could be a desirable property of a DNN. A convolutional recurrent neural network (CRNN) \cite{choi2016ConvolutionalRNN} was proposed, to this effect. It extracts local features using a 4-layer CNN frontend, summarizes them with a recurrent neural network (RNN) backend, and feeds the resulting output to a single FC layer to predict relevant tags. Each CNN layer contains a convolution with 2D $3 \times 3$ filters and a max-pooling operation. Different sizes of filters and strides are used for max-pooling layers ((2, 2), (3, 3), (4, 4), (4, 4)) to downsample the input before passing it to the RNN. It also uses 29.1s-long audio chunks (which are converted to mel-spectrograms) to perform song-level training. 
            
            CNN + SA \cite{won2019cnnsa}, is another, more recent, sequence modeling network. It leverages the self-attention operation \cite{cheng2016selfattention} to better model long-range dependencies and avoid the vanishing/exploding gradient problem that occurs in RNNs. It contains a CNN frontend composed of a deep stack of seven 2D convolutional blocks. Each block integrates a convolutional operation with $3 \times 3$ 2D filters, a residual connection, and a max-pooling operation. Long-term dependencies are learned by feeding the extracted features from the frontend to a backend network. It is composed of BERT (short for Bidirectional Encoder from Transformers) \cite{Devlin2019BERT}, which is a very popular architecture for various NLP tasks. BERT uses a deep stack of self-attention layers. CNN + SA uses 15s-long audio excerpts (which are converted to mel-spectrograms) to perform chunk-level training. 
            
            We note that the main differences from previous sequential networks (CRNN and CNN + SA) and our proposed models are that ours operate directly on waveform inputs, integrate multi-scale and multi-level features, attend jointly to temporal and channel subspaces, and do not remove information from feature maps.

       \subsubsection{Waveform-base networks}\label{subsec:wavmodels}
            A small collection of recent studies experiment with models that can operate directly on raw waveform input representations \cite{dieleman2014end, lee2018samplecnnAppliedSciences,  kim2018samplelevelcnnIEEE}. We refer these models as ``waveform-based.'' The advantages of waveform-based models are that they can operate in an end-to-end, assumption-free fashion and can learn task/domain-specific filters, since they avoid preprocessing audio inputs and relax the time-frequency resolution trade-off implicit in spectrogram-based networks \cite{kimComparisonSampleCNNIEEE,zhu2016LearningMultiScale}. The main disadvantage is that they generally require more training data \cite{Nam2019DeepForMusicMetaReview, won2020eval}. 
            
            SampleCNN \cite{lee2018samplecnnAppliedSciences} was the first waveform-based CNN architecture to achieve competitive performance for music tagging. It is an end-to-end, assumption-free model. It contains a deep stack of ten 1D convolutional blocks. Each block consists of a convolutional operation with sample-level $1 \times 3$ 1D filters followed by a $1 \times 3$ max-pooling operation. However, the first block uses a stride of 3 to downsample the input. SampleCNN + SE \cite{kim2018samplelevelcnnIEEE} subsequently enhanced the representational power of SampleCNN by adding a residual connection \cite{he2016ResNet} and the squeeze-and-excitation (SE) \cite{Hu2018SqueezeandExcitationN} channel-wise attention operation to each convolutional block. Both sample-level networks also integrate multi-level feature modeling by applying channel-wise global max-pooling to the top three layers. The extracted channel-wise statistics from each layer are then concatenated along the channel dimension and fed to two FC blocks to predict relevant tags. A variant of SampleCNN was proposed to target multi-scale features in addition to multi-level ones. We call this variant ``SampleCNN (x9)'' \cite{lee2018samplecnnAppliedSciences}. It consists of a frontend-backend architecture. The frontend contains an ensemble of nine pretrained SampleCNNs. While the backend is a 2-layer fully connected network. Each SampleCNN in the frontend is trained in a supervised fashion on a different timescale by adjusting its filter and stride size in the first layer. Once trained, multi-scale and level features are extracted and summarized by a combination of average and max-pooling. The resulting outputs are then concatenated and feed to the backend to predict relevant tags. It's relevant to notice that SampleCNN (x9) is not an end-to-end network since it requires pretraining, unlike SampleCNN and SampleCNN + SE. SampleCNN, SampleCNN + SE, and SampleCNN (x9) use short audio clips (3-5s) for chunk-level training. Studies show that SampleCNN and SampleCNN + SE also perform well across different audio tasks including speech and acoustic scene recognition \cite{kimComparisonSampleCNNIEEE}. The main difference from other waveform-based models and the above three networks is that they use very small 1D filters (between 2 and 5) in all layers.


        To summarize, at a high level, the architectures described in sections \ref{subsec:specmodels} and \ref{subsec:wavmodels} can be distinguished by two factors: the input representation and the amount of music domain knowledge used in the design process. Previous studies tend to use spectrogram or one of its variants like mel-spectrogram as the input representation for a network \cite{Choi2016AutoTaggingFCN,pons2017end, lee2017multilevelmusictagging, won2019cnnsa, won2020harmonic}. While a small selection of recent studies explore raw waveform audio as the input representation for a network \cite{dieleman2014end,lee2018samplecnnAppliedSciences, kim2018samplelevelcnnIEEE}. Many previous studies incorporate music domain knowledge in the network design process on some level. This might involve designing filters to target harmonic information \cite{pons2017end,won2020harmonic}, using sequential methods to model long-range interactions \cite{choi2016ConvolutionalRNN,won2019cnnsa}, and/or integrating multi-scale and/or multi-level features to account for hierarchical dependencies of music tags \cite{lee2017multilevelmusictagging,pons2017end,kim2018samplelevelcnnIEEE}, while only a few studies use little or no music domain knowledge, or at least not explicitly \cite{lee2018samplecnnAppliedSciences, won2020eval}.

    \subsection{Attention and Transformers}\label{sec:attention_mechanisms}
        Attention mechanisms have enjoyed widespread adoption as a computational module for sequence modeling \cite{Bahdanau2015NeuralMTattention, vaswani2017AttentionIA, Devlin2019BERT, Bello2019AttentionAC}.  Notably, Bahdanau proposed integrating attention with a recurrent neural network (RNN) \cite{Bahdanau2015NeuralMTattention} for machine translation. Vaswani \textit{et al.} \cite{vaswani2017AttentionIA} introduced the Transformer architecture, obtaining state-of-the-art performance on machine translation, and subsequently, many other NLP tasks. The Transformer consists of an encoder and decoder, which both use a deep stack of self-attention \cite{cheng2016selfattention} and point-wise FC layers. It also incorporates positional information by augmenting inputs with absolute position embeddings. The position embeddings can be fixed or learned during training to model the dependency between elements at different positions in the input sequence. Devlin \textit{et al.} \cite{Devlin2019BERT} enhanced the Transformer by introducing BERT, or Bidirectional Encoder Representations from Transformers, which only uses the encoder part from the Transformer. BERT's main contribution is incorporating bidirectional information into the encoder by using a masked language modeling (MLM) pretraining task.  
        
        The primary component of the Transformer and its variants like BERT is self-attention \cite{cheng2016selfattention}. Self-attention is a form of attention that processes a sequence by replacing each element by a weighted average of the rest of the sequence. It also does not suffer from the vanishing/exploding gradient problem that is common in other sequential modeling techniques such as RNNs.
        
        
        Despite the inherent sequential structure of music, attention has not been widely explored in MIR, particularly for discriminative music tasks. Notable examples of attention in MIR include: \cite{huang2020popmusictransformer} and \cite{choi2019EncodingMSTransformerAttention} for music generation, \cite{liu2020sourceseparationattention} for source separation, \cite{kimComparisonSampleCNNIEEE,wang2019multirepresentationattention} for music tagging, and \cite{Park2019AttentionChord} for chord recognition. Previous studies on attention mechanisms for music tagging tasks focus on recalibrating convolutional features by addition \cite{kimComparisonSampleCNNIEEE} or gating \cite{wang2019multirepresentationattention}, and typically only attend to the channel subspace as opposed to the temporal subspace.

\section{Method}

    \begin{figure*}[!ht]
    \centering
    \includegraphics[width=1\textwidth]{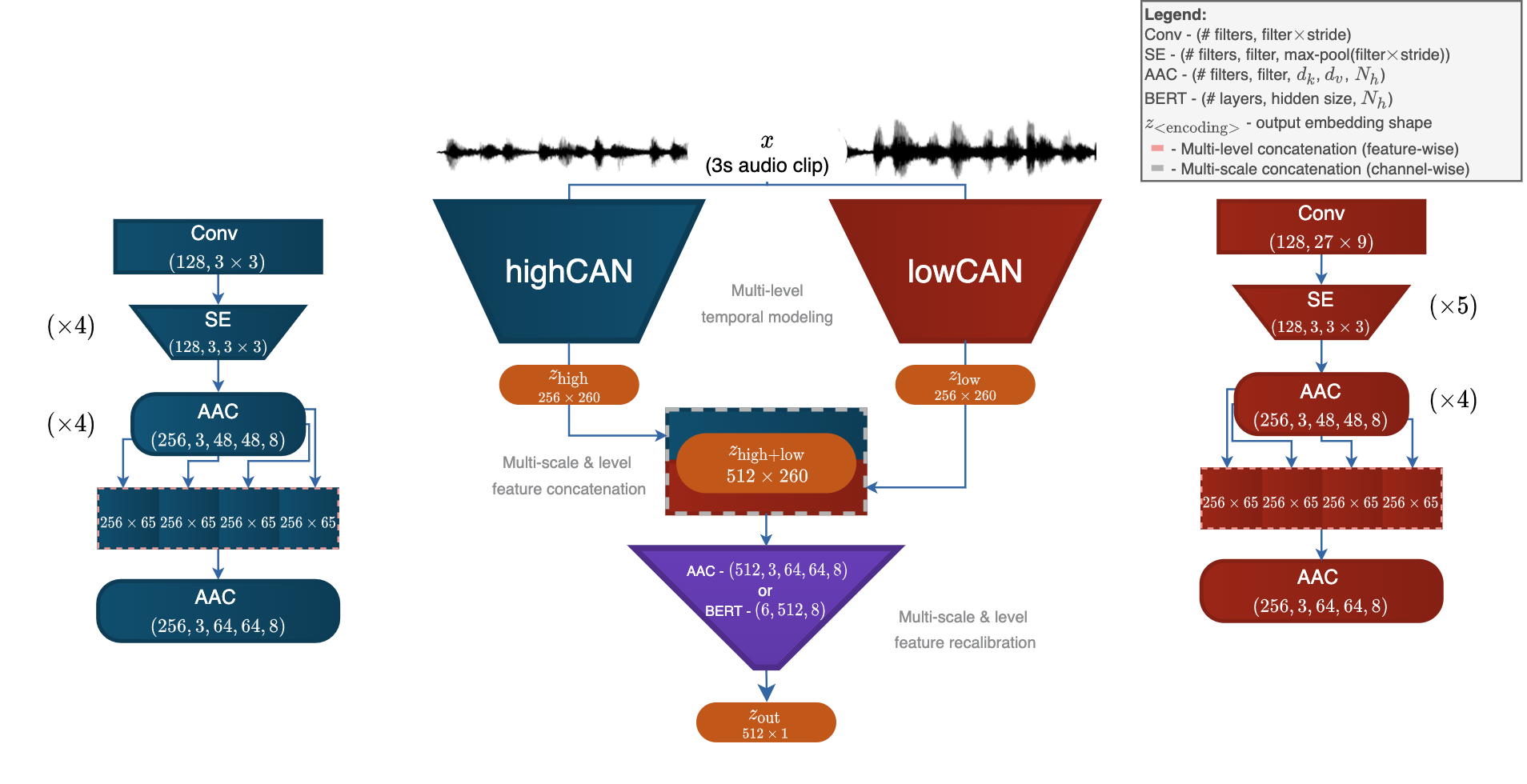}
    \caption{\textbf{Center}: The MuSLCAN/MuSLCAT architecture. \textbf{Left}: The small filter and stride in the first layer encourage highCAN to model high frequencies. \textbf{Right:} The relatively large filter and stride in the first layer encourage lowCAN to model low to mid frequencies. The high and low CANs use SE with max-pooling to recalibrate (via channel-wise statistics) and downsample feature maps. While AAC is used to recalibrate features and model long-term interactions by \textit{jointly} attending to both temporal and channel subspaces. The multi-level ouput embeddings from both CANs are channel-wise concatenated creating multi-scale and level feature maps, which are then recalibrated by either AAC or BERT.}
    \label{figure:dnn}
    \end{figure*}
    
    \label{sec:method} 
    In this section, we introduce the motivation behind our proposed waveform-based Multi-Scale, Multi-Level Convolutional Attention Transformer (MuSLCAT) architecture and formally detail its construction. We provide a high-level representation of MuSLCAT's structure in Figure \ref{figure:dnn}.
    
    \subsection{Motivation: Learning representations for complex music tags}
        Previous MAT / MGR studies suggest that individual music tags have different performance sensitivity to different timescales and levels of features \cite{lee2017multilevelmusictagging, lee2009unsupervisedmusicfeaturelearning, hamel2010multiscalelearning, dieleman2013multiscalemusictagging,sturm2014musicgenreinMIR,Lamere2008SocialTaggingAndGenreChallenges}. Unsurprisingly, most of the top-performing MAT / MGR networks enhance their representational power by combining features from different scales, levels, or both \cite{kim2018samplelevelcnnIEEE,Lee2017MultiLevelMultiScaleWaveSampleCNN,lee2017multilevelmusictagging}. In particular, two approaches are common. Most previous studies learn multi-scale and level features by using a three-stage training method: local feature learning, feature aggregation, and global classification \cite{Lee2017MultiLevelMultiScaleWaveSampleCNN,lee2017multilevelmusictagging}. In stage one, an ensemble of CNNs is trained in a supervised manner with tag labels, where each CNN takes different sizes of input. The feature aggregation stage extracts multiple-level features using the pre-trained CNNs and summarizes them into a single song-level feature vector. Finally, the last step uses a multi-layer fully-connected (FC) network to predict relevant tags. Another more recent approach is to only model multi-level features (but not multi-scale features) by using squeeze-and-excitation attention and multi-level feature aggregation \cite{kim2018samplelevelcnnIEEE}. The advantage being that it can be trained in a single stage (i.e., no pretraining). 
        
        \subsection{The MuSLCAT Architecture}\label{subsec:muslcat}
            Previous attempts to model multi-scale, multi-level, or multi-scale and multi-level features either sacrifice training simplicity for enhanced representational power (i.e., multi-stage training) or vice versa (i.e., multi-level modeling only), which is not ideal. Additionally, few previous studies model long-range interactions by learning directly from raw waveform audio. Thus, we present MuSLCAT, or Multi-scale and Multi-level Convolutional Attention Transformer, a novel architecture for learning robust representations of complex music tags directly from raw waveform recordings in an end-to-end fashion. We also introduce a lightweight variant of MuSLCAT called MuSLCAN, short for Multi-scale and Multi-level Convolutional Attention Network. Both MuSLCAT and MuSLCAN model features from multiple scales and levels by integrating a frontend-backend architecture. The frontend targets different frequency ranges while modeling long-range dependencies and multi-level interactions by using two convolutional attention networks with attention augmented convolution (AAC) blocks. The backend dynamically recalibrates multi-scale and level features extracted from the frontend by incorporating self-attention. The difference between MuSLCAT and MuSLCAN is their backend components. MuSLCAT's backed is a modified version of BERT \cite{Devlin2019BERT}. While MuSLCAN's is a simple AAC block. 
        
        \subsubsection{The Frontend}
            MuSLCAT's frontend combines two convolutional attention network (CAN) branches to efficiently learn useful representation from raw audio data. Each CAN captures temporal and hierarchical interaction using a mixture of convolution, squeeze-and-excitation (SE), and attention augmented convolution (AAC) blocks. We apply batch normalization \cite{ioffe2015batchnorm} and ReLU activation in convolution and SE blocks and layer normalization \cite{ba2016layernorm} in AAC blocks. We occasionally employ max pooling with filter and stride size 3 after blocks to reduce the network's memory footprint. Finally, we set the depth (i.e., number of layers) of the CANs to 10 and 11, respectively. 
        
            We encourage the CANs to emphasize different frequencies (e.g., low, mid, high) by adjusting their filter and stride size in the first layer. One CAN emphasizes high frequencies using a convolution with a $3 \times 3$ 1D filter. While the other CAN targets low to mid-level frequencies using a convolution with a $27 \times 9$ 1D filter. We name these two branches highCAN (red branch in Figure \ref{figure:dnn}) and lowCAN (blue branch in Figure \ref{figure:dnn}) according to the frequencies they aim to represent. We model multi-level feature interactions by concatenating feature maps from the top four layers and reweighing them using an AAC block in each CAN respectively.

        \subsubsection{The Backend}
             MuSLCAT's backend consists of two modules 1) our modified BERT architecture (detailed in section \ref{subsec:Transformer}), which recalibrate multi-scale and level features extracted from the frontend, and 2) a simple 2-layer FC classifier to predict relevant music tags. We apply the channel-wise concatenate procedure described in section \ref{subsec:multi-scale_features} to extract and combine multi-scale and level features from the frontend and recalibrate them using BERT. For BERT, we use a 6-layer 512-feature model with 8 attention heads. We set the dropout rate to 0.2 in self-attention and feed-forward operations respectively. We treat the multi-scale and level features as BERT's input embeddings, and add a [CLS] token to the beginning of each sequence as described in \cite{Devlin2019BERT}, since our goal is multi-label binary classification. The activations of the highest layer of the model at the [CLS] token are treated as the final feature representation of the multi-scale and level audio features. These features are then linearly projected to the prediction space using the 2-layer FC classifier. 
    
    \subsection{Attending to important features}
        Here we describe the attention mechanisms used in MuSLCAT.
          
        \subsubsection{Squeeze-and-Excitation Attention}
            The convolution operation is limited by its locality and lack of understanding of global context, and multiple MIR studies suggest methods to overcome these limitations \cite{lee2017multilevelmusictagging,lee2018samplecnnAppliedSciences,kim2018samplelevelcnnIEEE}. In particular, SampleCNN + SE \cite{kim2018samplelevelcnnIEEE} extends the original SampleCNN \cite{lee2018samplecnnAppliedSciences} by integrating squeeze-and-excitation (SE) \cite{Hu2018SqueezeandExcitationN} blocks, resulting in state of the art performance for waveform-based networks. 
          
            The SE attention method performs channel-wise recalibration of convolutional feature maps via two operations. First, a \textit{squeeze} operation applies global average pooling along the temporal dimension of feature maps to produce channel-wise statistics, resulting in a $(C \times 1)$ dimensional tensor (we omit the batch dimension for clarity), where $C$ represents the number of filters in the convolution. Then, an \textit{excitation} operation feeds the squeeze operation's output to a simple gating mechanism to produce per-channel modulation weights. The gating mechanism consists of two fully-connected (FC) layers that compute nonlinear interactions among channels. Finally, channel-wise multiplication between the original convolution's output and the excitation operation's output recalibrates the feature maps, producing the final feature maps.
            
        \subsubsection{Self-Attention over waveforms}\label{subsec:self-attention}
            Introduced in \cite{vaswani2017AttentionIA} for NLP tasks, self-attention is a recent advance to capture long-range interactions in sequential data. It works by computing a weighted average of values from hidden units using a similarity function that can dynamically adjust the weights. The use of multi-head attention in self-attention enables it to attend \textit{jointly} to both temporal and channel subspaces and remove none of the temporal information, unlike SE, which only attends to channels and removes temporal information via a global aggregation operation. 
            
            Below we detail self-attention over waveforms. We use the following naming conventions: $L$ and $C_{in}$ refer to the sequence length and number of input channels of a feature map. $N_h$, $d_v$ and $d_k$, $h$ respectively refer to the number of heads, the depth of values, the depth of queries and keys, and a single-head in multi-head attention.  Additionally, we omit the batch dimension for clarity.
         
            Given an input tensor $X$ of shape $(L, C_{in})$, we compute multi-head attention (MHA) as proposed in the original Transformer architecture \cite{vaswani2017AttentionIA}. The output of the self-attention mechanism for a single head $h$ can be formulated as: 
            \begin{equation}
                O_{h} = \text{Softmax}\left( \frac{ \left(XW_{q}\right) \left(XW_{k}\right)^T }{ \sqrt{d^h_k} } \right) \left(XW_{v}\right),
            \end{equation}
            where $W_{q}$, $W_{k} \in \mathbb{R}^{C_{in} \times d_{k}^h}$, and $W_{v} \in \mathbb{R}^{C_{in} \times d_{v}^h}$ are learned linear transformations that projects the input $X$ to queries $Q = XW_{q}$, keys $K = XW_{k}$ and values $V = XW_{v}$. The outputs of all heads are then concatenated and projected again as follows:
            \begin{equation}
                \text{MHA}\left(X\right) = \text{Concat}\left[O_{1},...,O_{N_{h}}\right]W^O ,
            \end{equation}
            where $W^O \in \mathbb{R}^{d_{v} \times d_{v}}$ is a learned linear transformation. $MHA(X)$ is then reshaped into a tensor of shape $(L, d_v)$ to match the original temporal dimension. The MHA operation incurs a complexity of $O(L^2d_k)$ and a memory cost of $O(L^2N_h)$, since it stores attention maps for each head. 
            
            \textbf{Positional Embeddings}: Without explicit information about positions, self-attention is \textit{permutation equivariant} \cite{Bello2019AttentionAC, shaw2018relativeselfattention}, which may not be optimal for modeling highly structured data such as music. Several studies suggest positional encoding methods that augment feature maps with explicit temporal information to address permutation issues. In particular, the original Transformer \cite{vaswani2017AttentionIA} introduces absolute position representations, using either positional sinusoids or learned position embeddings that are added to the per-position input representations. 
            
            However, absolute positional encodings do not satisfy \textit{translation equivariance}, which we hypothesise is a desirable property for processing music data. The work of Shaw \textit{et al.} \cite{shaw2018relativeselfattention} introduces relative position representations for the original Transformer, which allows attention to be informed by how far two positions are apart in a sequence. This involves learning an additional relative position embedding $E^r$ of shape $(N_h, L, d^h_k)$, which has an embedding for each possible pairwise distance $r = j_k - i_q$ between a query and key in position $i_q$ (the i-th row of $Q$) and $j_k$ (the j-th row of $K$) respectively. The embeddings are learned separately for each head. The self-attention output for head $h$ now becomes:
            \begin{equation}
                O_{h} = \text{Softmax}\left( \frac{QK^T + S^{rel}}{\sqrt{d^h_k}}\right)V,
            \end{equation}
            where $S^{rel} \in L \times L$ is the matrix of relative position logits that satisfies $S^{rel}[i,j] = Q_i^Tr_{j_k - i_q}$.
            
            The original implementation of relative attention \cite{shaw2018relativeselfattention} explicitly stores all relative embeddings $r$ in a tensor of shape $(L, L, d^h_k)$, which incurs an additional memory cost of $O(L^2d^h_k)$, thus restricting its application to long sequences and large batch sizes. The work of \cite{huang2018memoryefficientmusictransformer} introduce a memory efficient version of relative masked attention. We slightly modify this approach to unmasked relative self-attention for music audio recordings, thus reducing the relative positional embedding memory cost from $O(L^2d^h_k)$ to $O(Ld^h_k)$.
            
        \subsubsection{Attention Augmented Convolution}
            Introduced in \cite{Bello2019AttentionAC} for image recognition and object detection tasks, attention augmented convolution (AAC) augments convolutional feature maps with self-attention features maps. This is achieved by concatenating convolutional features and MHA features along the channel axis. The combination of a convolution operation and a self-attention operation enables an AAC block to attend \textit{jointly} to both channel and temporal subspaces, and add additional features to convolutional feature maps rather than refining them like in an SE block. It also makes it easy to adjust the proportion of attentional versus convolutional channels in each block. These properties make AAC an ideal candidate for our task, hence we modify the original implementation to operate on 1D data (i.e., raw waveforms) and integrate it in our proposed MuSLCAT network. Formally, given a convolution with input channels and output channels represented by $C_{in}$ and $C_{out}$ respectively, we can describe the corresponding AAC as:
    
            \begin{equation}
                \text{AAConv}\left(X\right) = \text{Concat}\left[\text{Conv}\left(X\right), \text{MHA}\left(X\right)\right],
            \end{equation}
            
            where $v = \frac{d_{v}}{C_{out}}$ denotes the ratio of attentional channels to the number of original convolutional channels and $k = \frac{d_{k}}{C_{out}}$ denotes the ratio of key depth to the number of output channels. 
            In terms of AAC's impact on the number of parameters, the MHA attention component introduces a $1\times1$ convolution with $C_{in}$ input filters and $(2d_k + d_v) = C_{out}(2k+v)$ output filters to compute query, key, and value features, and an additional $1\times1$ convolution with $d_v = C_{out}(v)$ input and output filters to mix the contributions of all the heads. Thus, the number of parameters can then be approximated by
            \begin{equation}
            C_{in}C_{out}(2k + (1 - r^2)v + \frac{C_{out}}{C_{in}}v^2),
            \end{equation}
            where $r$ denotes the original convolution operation's kernel size\footnote{We ignore relative position embedding parameters for simplicity since these are negligible.}. In practice, this yields a slight reduction in parameters when replacing $r = 3$ convolutions and a slight increase in parameters when replacing $r = 1$ convolutions. Interestingly, in experiments we find that our AAC enhanced networks achieve competitive results and require fewer parameters than state-of-the-result waveform-based models.
            
    
    \subsection{Multi-level feature modeling}\label{subsec:multi-level_features}
        Multiple studies suggest music tags have different levels of abstraction, and so combining features from different levels improves performance \cite{lee2017multilevelmusictagging, lee2009unsupervisedmusicfeaturelearning, hamel2010multiscalelearning, dieleman2013multiscalemusictagging, kim2018samplelevelcnnIEEE, Lee2017MultiLevelMultiScaleWaveSampleCNN}. The common multi-level feature modeling method has two steps: extract multi-level features and summarize them into a single feature vector. The first step extracts features from the last $l$ layers ($l \leq d$). The next step applies global pooling (either average or max) to produce channel-wise statistics for each layer, resulting in tensors of shape $(C_l \times 1)$, where $C_l$ is the number of convolutional filters in the $l$-th layer\footnote{We ignore the batch dimension for clarity.}. The summarized per layer features are finally concatenated to produce the final multi-level feature representations, which is a tensor of shape $(C \times 1)$, where $C$ equals the sum of all filters from the last $l$ layers.
        
        This multi-level approach removes potentially valuable information from the final multi-level feature representations by applying global average or max-pooling. We hypothesize this could limit the representational power of a multi-level model. As a solution, we suggest modeling the interactions between levels by concatenating features from different levels along the temporal axis and using an AAC block to recalibrate them instead. Formally, given a network with $d$ layers, multi-level depth of $l$ layers ($l \leq d$) and an AAC with $C_{out}$ output filters, our multi-level method can be formulated as:
        \begin{equation}
        O_{ml} = \text{AAConv}\left(\text{Concat}\left[O_{i},...,O_{d}\right]\right),
        \end{equation}
        where $O_{ml} \in \mathbb{R}^{C_{out} \times L_{out}}$ ($L_{out}$ is equal to the sum of the number of temporal features from all $l$ layers) and $O_{i}$ represents the features from the $l_i$-th layer ($d-l \leq i \leq d$). Note that our approach requires the last $l$ layers to have the same number of output filters. 
        
        
    
    \subsection{Multi-scale feature modeling}\label{subsec:multi-scale_features}
        For a CNN waveform-based MAT / MGR system to be useful, the filter bank in the first convolutional layer needs to learn representations that cover the entire frequency spectrum. But the fixed filter and stride size in the first convolutional layer tend to represent some frequencies better than others, with smaller filter and stride size emphasizing high frequencies  \cite{dieleman2014end,lee2018samplecnnAppliedSciences,zhu2016LearningMultiScale}. This property makes it difficult for a network to span the full frequency spectrum. A few multi-scale feature modeling approaches have been proposed to address this challenge. In particular, Zhu et al. \cite{zhu2016LearningMultiScale} introduce a frontend architecture for speech recognition that learns multi-scale representations by concatenating pooled feature maps from three separate convolutions. Each convolution employs a different filter and stride size to emphasize different frequencies (e.g., low, mid, or high). Then, a backend network passes the aggregated multi-scale features through several layers and makes predictions. Lee et al. develop a similar method for music tagging \cite{lee2017multilevelmusictagging,lee2018samplecnnAppliedSciences}. However, instead of an end-to-end network, an ensemble of pretrained deep CNNs (trained via supervision on the target task) is used to extract features from different scales. A combination of average and max-pooling is applied to the extracted features yielding channel-wise summary statistics, which are then concatenated and feed to a 2-layer FC network to predict relevant tags.
        
        Inspired by these approaches, we also integrate a multi-scale method in MuSLCAT. It involves two convolution attention networks and feature concatenation. Different filter and stride sizes are used in the convolutional attention networks to target different scales. The features are extracted from the convolutional attention networks and then concatenated along the channel axis yielding a multi-scale feature representation. Importantly, we do not summarize multi-scale features before concatenating them. 
        
        Formally, given two CAN branches with $C_{low}$ and $C_{high}$ output filters and $L_{low}$ and $L_{high}$ output features respectively, our multi-scale method can be formulated as:
        \begin{equation}
            O_{ms} = \text{Concat}\left[O_{low}, O_{high}\right]
        \end{equation}
        where $O_{ms} \in \mathbb{R}^{(C_{low} + C_{high}) \times (L_{low} + L_{high})}$. The main differences from previous multi-scale methods and our method is that ours 1) trains end-to-end (i.e., no pretraining) and 2) recalibrates concatenated feature maps instead of pooling them (i.e., no information is lost).
    
    \subsection{BERT}\label{subsec:Transformer}
          We implement BERT \cite{Devlin2019BERT} in MuSLCAT's backend to model interactions between multi-scale and level features. We closely follow BERT's original implementation with only slight modifications: we remove the global absolute positional embeddings and add memory efficient unmasked relative position embeddings to each self-attention layer (as described in section \ref{subsec:self-attention}), and we use the multi-scale and level feature embeddings extracted from the frontend as the input embeddings. We also do not pretrain BERT.
    
\section{Experimental Setup}\label{sec:experimentsetup}
    \begin{table}[!ht]
    \renewcommand{\arraystretch}{1.2}
    \caption{Music auto-tagging and genre recognition datasets used in our experiments.}
    \label{tab:datasets}
    \resizebox{0.95\linewidth}{!}{%
        \begin{tabular}{@{}lcccc@{}}
        \toprule
        Dataset            & MTAT         & MuMu                  & FMA               & MTG-Jamendo \\ \midrule
        Main Task          & Autotagging  & Genre                 & Genre             & Autotagging \\
        Size (days)        & $\sim$9      & $\sim$23$^2$          & $\sim$343         & $\sim$157   \\
        Tags$^1$           & 188 (50)     & 188 (188)             & 161 (154)         & 195 (50)    \\
        Annot. source      & Crowdsourced & Taxonomy$^3$          & Artist            & Artist      \\
        Annot. level       & Track        & Album                 & Track             & Track       \\
        Clips              & $\sim$25.9k  & $\sim$66k$^2$         & $\sim$106.6k      & $\sim$55.6k  \\
        Tracks             & $\sim$5.4k   & $\sim$66k$^2$         & $\sim$106.6k      & $\sim$55.6k  \\
        Artists            & 230          & N/A                   & $\sim$16.3k       & $\sim$3.6k  \\
        CC-licensed        & Yes          & No$^4$                & Yes               & Yes         \\
        Bitrate (Kbps)     & 32           & 104                   & 263               & 320         \\ \bottomrule
        \end{tabular}%
    }
    \\
    \tiny{$^1$ Numbers listed in parentheses show the total number of tags used in experiments. In all cases, the top-n most frequent tags were selected. $^2$ The original MuMu set contains $\sim$147k clips/tracks, but we were only able to retrieve audio previews for $\sim$66k clips/tracks. $^3$ Denotes the Amazon 4-level genre taxonomy. $^4$ 30-second audio previews for some tracks can be downloaded from streaming services.}
    \end{table}

    In this section, we detail the datasets used in our experiments (summarized in Table \ref{tab:datasets}) and describe experimental settings. 

    \subsection{Datasets}\label{sec:datasets}
        We evaluate the proposed MuSLCAT architecture on four popular MIR datasets MuMu \cite{oramas2017mumu} and FMA \cite{fma_dataset} for multi-label genre tagging and MTAT \cite{law2009mtat} and MTG-Jamendo \cite{bogdanov2019jamendo} for multi-label music tagging. We use an 80\%/10\%/10\% training, validation, testing split for the FMA and MuMu dataset. For MTG-Jamendo, we use split-0 provided by the set's creators\footnote{\label{note:mtg-jamendo} https://github.com/MTG/mtg-jamendo-dataset}. For MTAT, we follow the common split used by previous studies (see section \ref{subsec:MTAT} for details). In line with previous studies, all splits apply an artist filter to avoid the "artist effect" (e.g., when tracks by the same artist appear across sets) which can lead to over-optimistic model performance \cite{flexer2009albumartisteffects, sturm2014musicgenreinMIR, fma_dataset, lee2017multilevelmusictagging}. For all datasets, we resample audio recordings at 16kHz.
    
        \subsubsection{MuMu}\label{subsec:mumu}
            The MuMu dataset \cite{oramas2017mumu} includes genre annotation based on the Amazon 4-level genre taxonomy. It contains approximately 135k tracks from 31k albums, arranged in a hierarchical taxonomy of 446 genres. The genre annotations are provided at the album-level and extended to the track(s) associated with each album. We discard tags with fewer than 200 annotated tracks, which reduces the total genres to 188. Since MuMu does not provide audio for tracks, we use the Spotify API\footnote{https://developer.spotify.com/documentation/web-api/} to download 30 second audio previews for tracks. We note that the audio preview is not always available, and when it is, we don't have control over from which part in the song the preview is extracted (e.g., beginning, middle, or end). After cleaning and retrieving audio, our final MuMu dataset contains 188 genres and approximately 66k tracks. The advantage of MuMu is that it represents commercially popular music from a wide variety of genres on music streaming platforms. 
        
        \subsubsection{FMA}\label{subsec:fma}
            The FMA, or Free Music Archive, dataset \cite{fma_dataset} is an open and easily accessible large-scale dataset suitable for evaluating several tasks in MIR, genre recognition, in particular. It provides 917 GiB and 343 days of Creative Commons-licensed full-length audio for 106,574 tracks from 16,341 artists and 14,854 albums, arranged in a hierarchical taxonomy of 161 genres. We use the large FMA subset for our experiments, which contains the full dataset with audio limited to 30 seconds clips extracted from the middle of tracks (the entire track is used if it's shorter than 30 seconds). We also discard genre tags with fewer than 20 tracks. After cleaning, our final FMA dataset contains 154 genres and approximately 106k tracks.
        
        \subsubsection{MTAT}\label{subsec:MTAT}
            The MagnaTagATune (MTAT) \cite{law2009mtat} dataset is one of the most frequently used datasets for benchmarking automatic music tagging systems. It contains multi-label annotations by genre, mood, and instrumentation for 25,877 30-second long audio segments from 5,405 tracks and 230 artists. Typically, only the 50 most frequent tags are used for the task, which include tags for genre and instrumentation labels, decades (e.g., ’80s’ and ’90s’) and moods. The dataset is split into 16 folders, and it is common to use the first 12 folders for training, the 13th for validation, and the last three for testing. Thus, we follow the same data split, and use the top 50 tags to be consistent with results reported in previous studies \cite{won2020eval, lee2018samplecnnAppliedSciences, kim2018samplelevelcnnIEEE}\footnote{https://github.com/minzwon/sota-music-tagging-models}.
            
        \subsubsection{MTG-Jamendo Dataset}\label{subsec:jamendo}
            The MTG-Jamendo (Jamendo) \cite{bogdanov2019jamendo} is a newly available open and accessible large-scale music tagging dataset. It contains audio for 55,701 full songs and 195 different tags covering genres, instrumentation, moods and themes, and is built using music publicly available on the Jamendo\footnote{https://www.jamendo.com/} music platform under Creative Commons licenses. The minimum duration of each song is 30s, and they are provided in the MP3 (320Kbps bitrate) format. Overall, the average song duration is approximately 224s (3m 40s). In contrast to prior music tagging datasets, MTG-Jamendo contains significantly larger audio segments with higher encoding quality. This may be a result of the fact that Jamendo targets its business on royalty free music for commercial use, including music streaming for venues, and ensures a basic technical quality assessment for their collection. Thus, the audio quality level may be more consistent with commercial music streaming services, making it an ideal dataset for our purposes. The creators of the dataset provide multiple splits for training, validation and test. For all splits, steps were taken to ensure no track appears in more than one set and no tracks in any set are from the same artist present in other sets. Moreover, the same tags are present in all three sets across all splits, and tags are represented by at least 40 and 20 tracks from 10 and 5 artists in training and validation/testing sets, respectively. In this work we use split-0 and the 50 most frequent tags$^\text{\ref{note:mtg-jamendo}}$. 
            

    \subsection{Training/Evaluation Details}\label{subsec:train-eval-details}
        Previous music tagging studies demonstrate that models trained on short audio excerpts (3-5 seconds) outperform models trained on relatively longer excerpts (15+ seconds) \cite{won2020eval, lee2017multilevelmusictagging, kim2018samplelevelcnnIEEE}. Thus, we train our models on batches of randomly selected short audio clips (3s of raw waveform data). During training, we minimize binary cross-entropy loss, and update trainable parameters using stochastic gradient descent (SGD) with Nesterov momentum 0.9 and initial learning rate 0.01. We reduce the learning rate by a factor of 5 after model performance on the validation dataset plateaus for three consecutive epochs, and we stop training once the learning rate drops below $1.6 \times 10^{-5}$. We fix the batch size to 23, as in prior studies \cite{lee2018samplecnnAppliedSciences, kim2018samplelevelcnnIEEE} to fairly compare MuSLCAT to the state-of-the-art waveform-based music tagging networks. All models are trained on a single NVIDIA 2080ti GPU and implemented in the PyTorch \cite{Paszke2019PyTorchAI} deep learning framework. 
        
        Since training on larger datasets is more expensive, researchers typically prototype on a smaller dataset (e.g., MTAT or MuMu) and then validate their models on a larger dataset (e.g., FMA or Jamendo) \cite{bogdanov2019jamendo, lee2018samplecnnAppliedSciences, kim2018samplelevelcnnIEEE, won2020eval}. We adopt a similar approach by exploring different MuSLCAT configurations on MuMu, a medium size dataset, and then training and evaluating the most promising configuration on the three remaining datasets: FMA and Jamendo (large datasets) and MTAT (small dataset).
        
        We follow the approach of previous studies for evaluation \cite{lee2018samplecnnAppliedSciences, kim2018samplelevelcnnIEEE}. That is, for all models, we pass as input 3s long clips belonging to the same audio recording and average the model's outputs on these clips to produce a final song-level tag prediction. For metrics, we report Area Under the Receiver Operating Characteristic (ROC-AUC) curve and Area Under Precision-Recall (AUC-PR) curve, as the latter can be more informative for evaluation on imbalanced datasets \cite{davis2006aucpr}. We compute the average score across all tags for both metrics on each dataset respectively.

        \begin{table*}[!ht]
        \caption{Performance of MuSLCAT compared to state-of-the-art waveform-based models.}
        \label{tab:sotawaveform}
        \begin{adjustbox}{width=0.95\textwidth}
            \begin{tabular}{@{}lcccccccc@{}}
            \toprule
            \multicolumn{1}{c}{\multirow{2}{*}{Architecture}} & \multicolumn{2}{c}{MTAT}  & \multicolumn{2}{c}{MTG-Jamendo} & \multicolumn{2}{c}{MuMu} & \multicolumn{2}{c}{FMA} \\ \cmidrule(l){2-9} 
            \multicolumn{1}{c}{}                                       & ROC-AUC    & PR-AUC                & ROC-AUC   & PR-AUC                & ROC-AUC   & PR-AUC                & ROC-AUC   & PR-AUC \\ \midrule
            SampleCNN\cite{lee2018samplecnnAppliedSciences}            & 0.9058     & 0.4422                & 0.8208    & 0.2742                & 0.8759     & 0.2146               & 0.8294    & 0.1506 \\
            SampleCNNs (x9)*\cite{lee2018samplecnnAppliedSciences}     & 0.9064     & -                     & -         & -                     & -         & -                     & -         & -      \\
            SampleCNN + SE\cite{kim2018samplelevelcnnIEEE}             & \textbf{0.9103} & \textbf{0.452}   & 0.8233    & 0.2784                & 0.8789    & 0.2211                & 0.8270    & 0.1530 \\ \midrule
            MuSLCAN                                                    & 0.9076     & 0.4439                & 0.8212    & 0.2762                & 0.8757    & \textbf{0.2250}       & 0.8300    & \textbf{0.1543}  \\ 
            MuSLCAT                                                    & 0.9061     & 0.4367                & \textbf{0.8239}  & \textbf{0.2793} & \textbf{0.8798} & 0.2219         & \textbf{0.8356} & 0.1488  \\ \bottomrule  
            \end{tabular}%
        \end{adjustbox}
        \\
        \tiny{* Denotes that the model used an ensemble of nine pretrained CNNs.}
        \end{table*}
    
\section{Results \& Discussion}\label{sec:results-discussion}
    
    In this section, we present the results from our experiments and offer our interpretation of them. 
    
    \subsection{Comparison of MuSLCAT components}\label{subsec:compare-muslcat-components}
        \begin{table}[!ht]
        \renewcommand{\arraystretch}{1.3}
        \caption{Performance of MuSLCAT's different architectural components on MTAT.}
        \label{tab:muslcatcomparison}
        \resizebox{0.95\linewidth}{!}{%
        \begin{tabular}{@{}lccc@{}}
        \toprule
        Architecture   & Params (M) & ROC-AUC & PR-AUC \\ \midrule
        lowCAN (low)   & 1.12      & 0.9044  & 0.4322  \\
        low + BERT     & 14.65     & 0.9064  & 0.4366  \\
        highCAN (high) & 1.25      & 0.9053  & 0.4373  \\
        high + BERT    & 14.70     & 0.9058  & 0.4345  \\
        lowCNN + highCNN & 3.34    & 0.9056  & 0.4384  \\ \midrule
        MuSLCAN     & 3.38      & 0.9076  & 0.4439  \\
        MuSLCAT & 15.70  & 0.9061 & 0.4367 \\ \bottomrule
        \end{tabular}%
        }
        \\
        \tiny{MuSLCAN is composed of low + high CANs and an AAC block to recalibrate multi-scale and level features. MuSLCAT is composed of low + high CANs and BERT to recalibrate multi-scale and level features.}
        \end{table}

        We first investigate the effectiveness of different MuSLCAT components by comparing their performance on the small-scale MTAT dataset. Table \ref{tab:muslcatcomparison} summarizes the evaluation results. They confirm our intuition that combining multi-scale and level features improves performance. A surprising result is that AAC outperforms BERT as a multi-scale and level representation modeling technique for this small-scale dataset. Additionally, they show that relative self-attention (used in AAC blocks and in BERT) improves MuSLCAT's performance. 
        
        Regarding multi-level feature modeling, AAC, which recalibrates features without losing information, outperforms the popular global pooling approach. For multi-scale feature modeling, the combination of both lowCAN and highCAN features results in improved performance, even when a simple AAC block is used to recalibrate multi-scale features instead of our BERT backend. However, performance w.r.t. PR-AUC decreases without our BERT backend. This result may suggest that our BERT backend learns more robust representations for underrepresented tags than other approaches.
        
    
    \subsection{Comparison of MuSLCAT to state-of-the-art}
        \begin{table}[ht]
        \renewcommand{\arraystretch}{1.3}
        \caption{Performance of MuSLCAT compared to state-of-the-art spectrogram-based models. }
         \label{tab:sotaspec}
        \resizebox{0.95\linewidth}{!}{%
        \begin{tabular}{@{}lcccc@{}}
        \toprule
        \multicolumn{1}{c}{\multirow{2}{*}{Architecture}} & \multicolumn{2}{c}{MTAT} & \multicolumn{2}{c}{MTG-Jamendo} \\ \cmidrule(l){2-5} 
        \multicolumn{1}{c}{}    & ROC-AUC & PR-AUC & ROC-AUC & PR-AUC \\ \midrule
        CRNN \cite{choi2016ConvolutionalRNN}                    & 0.8722  & 0.3625 & 0.7978  & 0.2358 \\
        FCN  \cite{Choi2016AutoTaggingFCN}                   & 0.9005  & 0.4295 & 0.8255  & 0.2801 \\
        MusiCNN \cite{pons2017end}                 & 0.9106  & 0.4493 & 0.8226  & 0.2713 \\
        Spec-SampleCNNs (x3)$^1$ \cite{lee2017multilevelmusictagging}   & 0.9017  & -      & -       & -      \\
        Harmonic CNN \cite{won2020eval}            & \textbf{0.9127}  & 0.4611 & \textbf{0.8322}  & \textbf{0.2956} \\
        Short-chunk CNN + Res\cite{won2020eval}    & 0.9126  & \textbf{0.4614} & 0.8316  & 0.2951 \\ \midrule
        MuSLCAN                 & 0.9076  & 0.4439  &  0.8212 &  0.2762     \\    
        MuSLCAT                 & 0.9061  & 0.4367  &  0.8239 & 0.2793      \\ \bottomrule
        \end{tabular}%
        }
        \\
        \tiny{$^1$ Denotes that the model used an ensemble of 3 pretrained CNNs. We include Spec-SampleCNNs (x3) because it incorporates multi-scale and level features and was shown to perform well on a large-scale dataset \cite{lee2017multilevelmusictagging}. ** denotes Short-Chunk CNN with residual connections (Res) \cite{won2020eval}.}
        \end{table}
        
        \textbf{Comparison to waveform-based models}: We first investigate how MuSLCAT performs on several benchmark music tagging and genre recognition datasets compared to the state-of-the-art waveform-based models. Table \ref{tab:sotawaveform} shows that MuSLCAT leads to systematic improvements on both auto-tagging and genre recognition tasks across medium-scale and large-scale datasets over the baseline network (SampleCNN) and the top-performing network (SampleCNN + SE). The results also suggest that AAC can be an efficient alternative to the computationally expensive SE operation with minimal impact on performance. 
        
        \textbf{Comparison to spectrogram-based models}: Though the focus of this work is learning directly from raw waveform input, we also compare MuSLCAN's performance to state-of-the-art spectrogram-based models on two benchmark music tagging datasets: MTAT (a small-scale dataset) and MTG-Jamendo (a large-scale dataset). Table \ref{tab:sotaspec} shows that MuSLCAT yields competitive performance compared to state-of-the-art spectrogram-based models on large-scale music tagging datasets while avoiding using task specific hand-crafted features and input preprocessing. 
        
        For multi-scale and level modeling, Table \ref{tab:sotawaveform} shows that MuSLCAN, which only contains a simple AAC block in its backend, can improve performance over the standard global pooling method. While, MuSLCAT, which replaces AAC with BERT in its backend, yields the best overall performance on large-scale datasets, and still requires fewer parameters than SampleCNN + SE. These results support our hypothesis that a negative side effect of SE and global pooling is that some useful information is removed from the signal, and we show that AAC can be used to overcome this.  
        
        Perhaps surprising to see is the impact training dataset size has on MuSLCAT's performance. On the small-scale MTAT dataset, MuSLCAT yields competitive results compared to other waveform-based models. Yet, on medium to large-scale datasets, MuSLCAT generally outperform waveform-based models. We hypothesis this to be the result of replacing global max pooling with MHA, since MHA recalibrates features using a learned weighted average whereas global max-pooling simply reduces temporal features to one for each channel. The global max-pooling operation may force the network to attend to only the most significant quality in a given signal. Intuitively, this design choice would be helpful when training data is limited (i.e., small or medium-scale), but could become a constraint for training on large-scale datasets. This would help explain the performance gain of MuSLCAT on MTG-Jamendo and FMA respectively. We also anticipate that the implicit biases associated with MHA make MuSLCAT more flexible to the domain data, which can enhance its representational capacity at the expense of needing more training data. 
        
\section{Conclusion}\label{sec:conclusion}
    In this work, we present MuSLCAT, or Multi-scale and Multi-level Convolutional Attention Transformer, a novel architecture for learning robust representations of complex tags directly from raw waveform music recordings. We also introduce MuSLCAN, or Multi-scale and Multi-level Convolutional Attention Network, a lightweight alternative to MuSLCAT. Both MuSLCAT and MuSLCAN capture features from multiple scales and levels by integrating a frontend-backend architecture. The frontend targets different frequency ranges while modeling long-range dependencies and multi-level interactions by using two convolutional attention networks with AAC blocks. The backend dynamically recalibrates multi-scale and level features extracted from the frontend by using BERT in MuSLCAT, or AAC in MuSLCAN.
    
    We validate the effectiveness of MuSLCAT and MuSLCAN by comparing them to state-of-the-art networks across four benchmark music tagging and genre recognition datasets. The experiments show that compared to state-of-the-art waveform-based models both MuSLCAT and MuSLCAN consistently yield competitive performance and require fewer parameters. In particular, given a large-scale training dataset, MuSLCAT produces the best results and is more computationally efficient ($34.2$\% reduction in parameters) than the current state-of-the-art network (SampleCNN + SE). On the other hand, MuSLCAN achieves slightly lower performance but requires significantly fewer parameters ($86.6$\% reduction) than the current state-of-the-art network. The main contributions of MuSLCAT and MuSLCAN include:
    
    \begin{enumerate}
    \item Attending jointly to channel and temporal subspaces by incorporating AAC blocks.
    \item Emphasizing different frequencies by using two CAN branches. This relaxes the need for each branch to represent the entire frequency spectrum in their bottom layer.
    \item Minimizing information loss by replacing the global pooling and SE operations with MHA to recalibrate feature maps instead of summarizing them with channel-wise statistics.
    \item Modeling long-range dependencies and multi-scale and level feature interactions by integrating AAC and/or BERT. 
    \item Training in an end-to-end fashion (i.e., no pretraining or preprocessing).
    \item Improving computational efficiency while outperforming state-of-the-art waveform-based networks, with an approximate $34.2$\% (MuSLCAT) and $86.8$\% (MuSLCAN) reduction in parameters compared to SampleCNN + SE.
    \end{enumerate}

\bibliography{ISMIR2020template}

%
%
%
%

\end{document}